# Experimental and Theoretical Investigation into the Effect of the Electron Velocity Distribution on Chaotic Oscillations in an Electron Beam under Virtual Cathode Formation Conditions

Yu. A. Kalinin and A. E. Hramov

*Chernyshevsky State University, Saratov, 410012 Russia*
*e-mail: aeh@cas.ssu.runnet.ru*

**Abstract**—The effect of the electron transverse and longitudinal velocity spread at the entrance to the interaction space on wide-band chaotic oscillations in intense multiple-velocity beams is studied theoretically and numerically under the conditions of formation of a virtual cathode. It is found that an increase in the electron velocity spread causes chaotization of virtual cathode oscillations. An insight into physical processes taking place in a virtual-cathode multiple-velocity beam is gained by numerical simulation. The chaotization of the oscillations is shown to be associated with additional electron structures, which were separated out by constructing charged particle distribution functions.

## INTRODUCTION

Hot interest in a new class of powerful electron devices that emerged on the border between 1970s and 1980s and embraces devices using a virtual-cathode (VC) electron beam as an active medium [1–4] still persists [5–9]. Even in pioneering experiments and calculations, the complicated time-varying radiation pattern of VC oscillators was noted [10–14]. VC chaotic oscillations were investigated both theoretically and experimentally [6, 8, 9, 14–18]. The occurrence of and interaction between coherent space–time structures in an electron beam with a VC were studied in [17, 19–23].

Experiments with VC oscillators, as well as their visualization, pose grave difficulties because of the need to use intense relativistic electron beams with currents exceeding a limit vacuum (supercritical) current [24]. Experimental conditions for generating microwave oscillations with a VC may be loosened by using additional slow-wave structures where a nonstationary oscillating VC is formed through severe slowdown of the beam (an electron beam with a supercritical perveance) 27, 28].[1] In such a structure, the VC formation and chaotic wide-band signal generation may take place at low currents and electron densities in the electron beam, which makes it possible to perform detailed experimental investigation of processes in a VC beam by applying physical experimental techniques to microwave electronics [29]. Such VC-containing systems with slowdown may serve as sources of noise-like wide-band chaotic microwave signals of medium power.

It was shown [27, 28, 30–32] that, in VC systems with additional slowdown, wide-band noise-like oscillations with a typical bandwidth of more than an octave are observed at certain values of the decelerating potential and beam current. These oscillations are fairly regular: irregularity parameter $N = P_{max}/P_{min}$ (where $P_{max}$ and $P_{min}$ are, respectively, the maximal and minimal powers in the oscillation power spectrum), is small. Note that the works cited above consider only single-velocity beams of charged particles. However, of much interest is to see how an electron velocity spread in the beam (multiple-velocity beam generated by an electron gun) influences the VC chaotic oscillations.

In this work, we theoretically and experimentally study the effect of an electron velocity spread (multiple-velocity bean generated by an electron–optical system with a thermionic cathode) on the parameters of wide-band chaotic oscillations in a VC beam.

## 1. EXPERIMENTAL SETUP

Oscillations in a VC beam were studied with a diode-type setup schematically shown in Fig. 1a. Here,

---

[1] Note that oscillations in such a slow-wave structure are to an extent similar to Barkhausen–Kurz oscillations [25] and also to those in an electron wave generator with a slowing-down field [26].

an electron beam generated by an electron–optical system (EOS) is injected into a slowing-down space between two grid electrodes. The decelerating (slowing-down) field was produced by applying negative (relative to entrance (first) grid 5) potential $V_{dec}$ to the exit (second) grid 6.

Hot filament (thermionic cathode) 1 serves as an electron source. The electrons emitted from the filament had a considerable velocity spread (for details, see Section 2). The EOS formed axisymmetric convergent cylindrical electron beam 4 with a high electron velocity spread. The beam-accelerating voltage was 2.0 kV, the beam current at the exit from the EOS was varied from 50 to 100 mA depending on the filament voltage, the radius of the beam was $r_b = 4$ mm, grid spacing $L$ was 20 mm.

Having escaped from the EOS, the beam with an initial velocity spread falls into the intergrid (diode) space. Potential $V_0$ of the first grid equals anode potential $V_a$ (accelerating voltage), potential $V_{dec} = V_0 - \Delta V_{dec}$ of the second grid was varied from $V_{dec}$ ($\Delta V_{dec} = 0$, deceleration is absent, classical Pierce diode) to zero ($\Delta V_{dec} = V_0$, complete deceleration of the electron beam). Quantity $\Delta V_{dec}$ is the potential difference between the grids that produces a decelerating field in the diode space.

As decelerating potential difference $\Delta V_{dec}$ reaches critical value $[\Delta V_{dec}]_{cr}$, a VC arises in the system. Its oscillations in time and space modulate the electron beam, with some of the electrons reflecting from the VC back to the entrance grid. As a result, the oscillations become chaotic, with their shape and power depending on potential difference $\Delta V_{dec}$ between the grids.

To analyze the noise-like oscillations of the electron beam, we used a wide-band segment of helical slow-wave structure (HSWS) 7 terminated by an absorbing insert and energy output 8 [29]. After passing through the slow-wave structure, the electron beam was directed to collector 9. The beam modulated in velocity and density in the diode space excites the HSWS segment, the signal from which is analyzed by an SCh-60 spectrum analyzer with the 200 MHz–19 GHz bandwidth and an S1-74 high-frequency analyzer. For signal analysis, we also used high-$Q$ filters (the bandwidth 2–4 MHz) configured with an ÉPP-09 oscilloscope. In this way, we determined the noise power spectral density of the oscillations generated by the VC electron beam.

The experimental vacuum setup was dismountable and operated under continuous evacuation (the residual gas pressure is $10^{-7}$ Torr). Its exterior view is shown in Fig. 1b.

## 2. EXPERIMENTAL RESULTS

First, we studied the structure of the beam generated by the hot-filament gun in the plane the beam enters the space of interaction ($z_0$ in Fig. 1a). The velocity spectrum was measured using the method of nonstationary diagrams and a decelerating-field analyzer [29, 33].

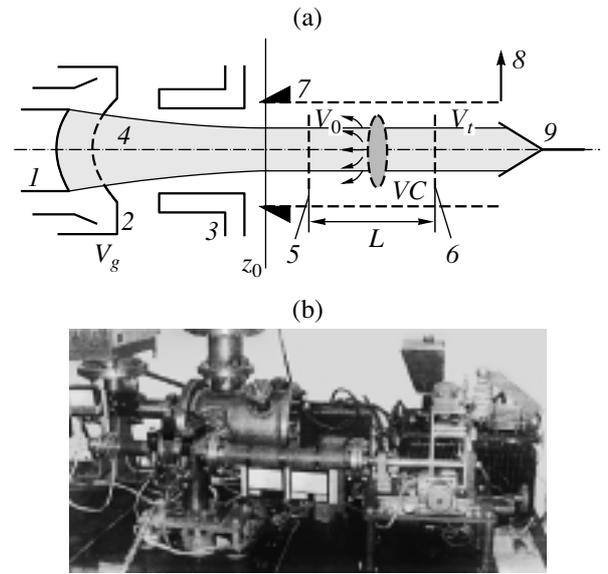

**Fig. 1.** (a) Schematic and (b) exterior view of the experimental setup used to study chaotic oscillations in a VC-containing beam with a supercritical perveance: (1) hot filament (filament voltage $V_f$), (2) grid of electron gun (grid potential $V_g$), (3) second anode of gun, (4) convergent electron beam, (5) entrance grid of diode space, (6) exit grid under decelerating potential, (7) HSWS segment, (8) energy output, and (9) collector.

Figure 2 shows the experimental electron distributions over angles, $f(\alpha)$, and longitudinal velocities, $f(v)$, at the entrance to the space of interaction for various operating conditions of the electron gun. In Fig. 2a, these distributions are shown for the electrons of the beam generated by the EOS in the space-charge-limited current regime (the filament current is $V_f = 6.3$ V, the grid is not specially biased, i.e., is under a "natural" potential). The measurements taken at the center of the beam ($r/v_b = 0.5$), are virtually independent of the radius. It follows from Fig. 2a that the spread of the angles of entry and longitudinal velocities us small: the half-widths of the angular and velocity distribution functions are $\Delta\alpha \sim 0.5$ and $\Delta v/v_0 \sim 0.25\%$, respectively. Such an operating regime is typical of the EOS of electron guns.

Other EOS operating conditions cause these distributions to broaden (i.e., multiple-velocity beams to form). The parameters of the electron beam formed at a high velocity spread (the gun operates under the temperature-limited current conditions, $V_f = 12$ V) and grid potential $V_g$ 1.6 times natural grid potential $V_{gn}$ are shown in Fig. 2a (the measurements were made at three points $r/v_b$ in the beam's cross section). In this case, the angular and longitudinal velocity distribution half-widths are $\Delta\alpha = 0.125$–$0.200$ and $\Delta v/v_0 = 0.2$–$2.0\%$

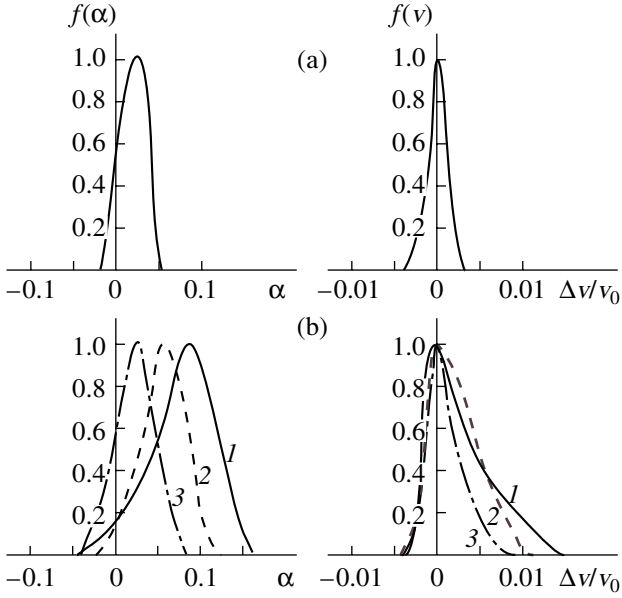

**Fig. 2.** Distribution of electrons over angles (on the left), $f(\alpha)$, and velocities (on the right), $f(v)$, at the entrance to the space of interaction. (a) Small velocity spread: the gun operates under the space-charge-limited current conditions (filament voltage $V_f = 6.3$ V) with the grid under the natural potential and measurements are taken at the center of the beam ($r/r_b = 0.5$). (b) Large velocity spread: the gun operates under the temperature-limited current conditions ($V_f = 12$ V); the grid potential is 1.6 times the natural potential; and measurements are taken at $r/r_b = $ (*1*) 0.9, (*2*) 0.5, and (*3*) 0 (three points in the beam's cross section).

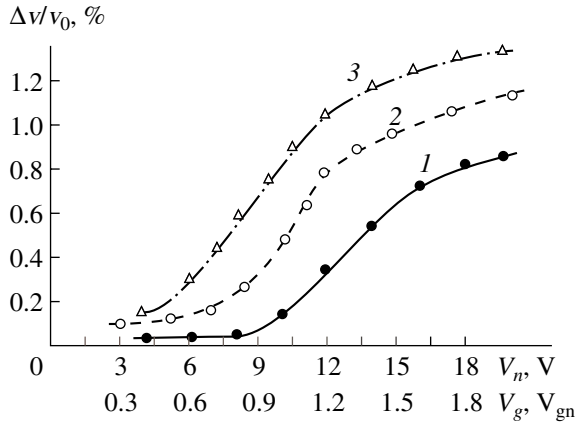

**Fig. 3.** Electron longitudinal velocity spread $\Delta v/v_0$ vs. (*1*) grid potential under the space-charge-limited current conditions ($V_f = 6.3$ V), (*2*) filament voltage with the grid under the natural potential, and (*3*) filament voltage with the grid potential 1.6 times the natural potential ($V_g = 1.6 V_{gn}$).

depending on the point of measurement in the cross section.

Figure 3 shows the variation of the longitudinal velocity spread with grid potential, filament potential, and both potentials simultaneously. The measurements were taken at point $r = 0.5 r_b$. The velocity spread of the electrons may be controlled in the range 0.05–2.50% by varying the filament voltage and the grid voltage relative to the natural value. Accordingly, one can trace the effect of the electron velocity initial spread on oscillations in the VC system.

Consider now how the oscillation parameters of the system vary with the electron velocity spread. As was mentioned above, oscillations in the system are caused by the emergence of a VC in the intergrid space with a decelerating field. Experiments and numerical simulations of the system with a decelerating field [27, 28, 31, 32] demonstrate that, at very low decelerating potential difference $\Delta V_{dec}/V_0$ between the grids, oscillations in the electron beam are absent. As the decelerating field (i.e., $\Delta V_{dec}/V_0$) grows and reaches critical value $[\Delta V_{dec}/V_0]_{cr}$, a VC arises in the system, which reflects some of the electrons back to the first grid. If the decelerating field is relatively low, VC oscillations are near-regular and the radiation spectrum is discrete. As the decelerating field increases further, wide-band noise-like oscillations occur. At a sufficiently high decelerating field, the oscillations cease.

Figure 4 illustrates the aforesaid for the system with a low electron velocity spread (single-velocity beam). In Fig. 4a, the oscillation normalized integral power is plotted against potential difference $\Delta V_{dec}/V_0$ between the grids at electron velocity spread $\Delta v/v_0 \sim 0.2\%$. Integral power $P_\Sigma$ was measured throughout the oscillation range and was normalized to the maximal integral power. It is seen that, when the decelerating field is low, so is integral power $P_\Sigma$ of oscillations in the VC beam. As $\Delta V_{dec}/V_0$ grows, the power increases, reaches a maximum at an optimal value of the decelerating field, and then drops. Figure 4a also plots current ratio $K$, which is defined as the time-averaged current passing through the output (second) grid, $\langle I_{out} \rangle$ ($\langle \rangle$ means averaging), to beam current $I_0$ at the exit from the electron gun; that is, $K = \langle I_{out} \rangle / I_0$. It follows from this figure that, as decelerating voltage difference $\Delta V_{dec}$ grows, the amount of the electrons reflecting from the VC or from the diode space downstream of the cathode back toward the first grid increases. When potential difference $\Delta V_{dec}$ between the grids becomes relatively high, the electrons totally reflect from the VC and so $K \approx 0$. Under these conditions, the VC in the electron beam becomes stationary and the oscillations cease (the integral power equals zero).

Figure 4b presents different typical oscillation conditions of the VC in the system with a single-velocity beam that correspond to different decelerating potential ranges. Shown are the ranges in which the VC oscillations are absent, regular, or wide-band chaotic. The steady state of the beam changes to chaotic oscillations through the periodic dynamics. As the decelerating potential grows, the chaotic oscillations become more complicated (the oscillation bands widen and the oscillations become more regular).

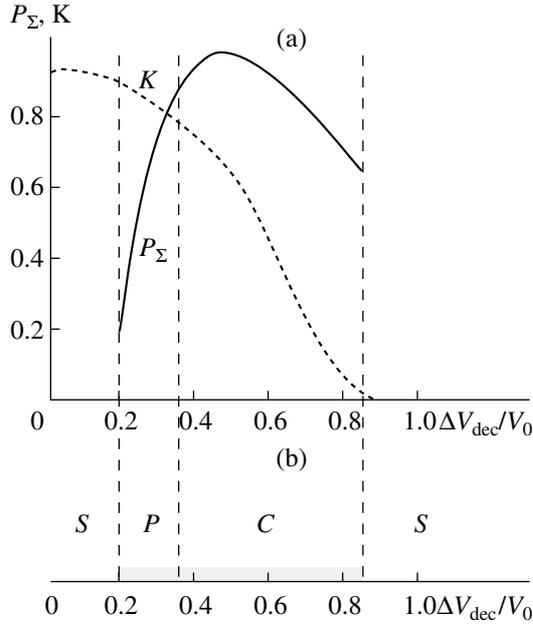

**Fig. 4.** (a) Normalized integral interaction power $P_\Sigma$ and current ratio $K$ vs. normalized potential difference $\Delta V_{dec}/V_0$ between the grids for the single-velocity beam and (b) characteristic oscillation regimes for the single-velocity VC beam: $A$, oscillations are absent; $R$, regular oscillations; and $C$, wide-band chaotic oscillations. The range where a nonstationary VC forms (microwave oscillations in our electron wave generator) is colored gray.

Let us consider the effect of the electron velocity spread with $\Delta V/V_0$ taken as a reflecting potential. In this case, wide-band chaotic oscillations are established in the system and the output integral power is close to a maximum value.

Figure 5a plots the variation of oscillation frequency $\Delta f/f$ and irregularity parameter $N$ of the oscillation power spectrum with grid voltage $V_g = V_{gn}$, filament voltage $V_f$, and both voltages simultaneously. The bandwidth and irregularity parameter describe the complexity of oscillations in the system. As the frequency band expands and the irregularity decreases, the oscillations become more and more chaotic and noise-like.

It is seen that bandwidth $\Delta f/f$ is maximal when both parameters are taken into consideration (curve $3$) and this maximum exceeds that for the single-velocity beam by a factor of 1.6. In the regime optimal for obtaining complicated chaotic oscillations with c continuous spectrum ($V_f = 20$ V, $V_g = 1.6 V_{gn}$), the irregularity decreases to 4 dB.

Note that oscillations in a VC beam become complicated with increasing velocity spread because of their dynamic nature (dynamic chaos) and not because of noise in the beam. Measurements showed that the mean noise intensity behind the second grid of the transit

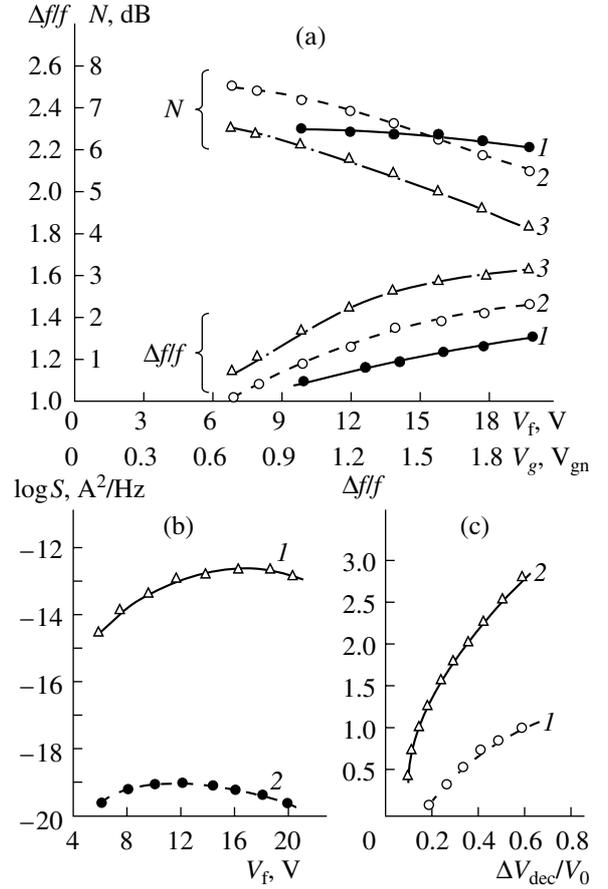

**Fig. 5.** (a) Oscillation frequency band $\Delta f/f$ and power spectrum irregularity $N$ for the VC beam vs. ($1$) grid potential (under the space-charge-limited current conditions, $V_f = 6.3$ V), ($2$) filament voltage (the grid under the natural potential, and ($3$) filament voltage with the grid potential 1.6 times the natural potential ($V_g = 1.6 V_{gn}$); (b) logarithm of noise mean intensity $S$ in the beam vs. filament voltage for ($1$) VC oscillations in the range 1–10 GHz and ($2$) shot noise; and (c) oscillation frequency band $\Delta f/f$ vs. decelerating potential for the ($1$) single-velocity beam and ($2$) multiple-velocity beam with $\Delta v/v_0 \approx 1.1\%$,

space is 60–70 dB higher than the mean intensity of shot noise, which equals $S_0 = 10^{-19}$ A$^2$/Hz. Figure 5b demonstrates mean noise intensity $S$ in the beam as a function of filament voltage $V_f$ for the system with $\Delta V_{dec}/V_0 = 0.5$ (shot noise versus $V_f$ according to [34] is also shown). The intensity of noise-like oscillations in the VC beam is much higher than the shot noise intensity, with the maxima of shot noise and noise-like oscillations observed at $V_f = 12$ and 18 V, respectively.

An important control parameter of the electron wave generator is the reflecting potential, with an increase of which the oscillations become more and more complicated, as was noted above. The dependence of the bandwidth of wide-band chaotic oscillations on the reflecting potential for the single-velocity beam and multiple-velocity beam with $\Delta v/v_0 \approx 1.1\%$ ($V_f = 12$ V and

$V_g/V_{gn} = 1.6$) is shown in Fig. 5c. In the latter case, the system starts oscillating at a lower decelerating potential, $\Delta V_t/V_0 \approx 0.05$, and the frequency band rapidly expands to an octave at $\Delta V_{dec}/V_0 \approx 0.2$. As the decelerating potential continues rising, the frequency band expands much more rapidly than for the single-velocity beam, reaching 2.5–3.0 octaves (for the single-velocity beam, the frequency band cannot cover more than an octave, see [31, 32]).

## 3. NUMERICAL SIMULATION OF PROCESSES IN THE MULTIPLE-VELOCITY VIRTUAL-CATHODE BEAM

Consider now the results of numerical simulation of nonlinear nonstationary processes taking place in the multiple-velocity VC beam and compare them with experimental data for chaotic oscillations. We will be interested in physical processes in the system with a high electron velocity spread.

We will use a 1D model of the transit space with a decelerating field and the large-particle method [35, 36]. Obviously, under certain operating conditions of the electron wave generator applied in this study, the electron flow cannot be viewed as one-dimensional. However, we may assume that basic physical processes responsible for the VC formation in the diode space are the same in both 1D and more complicated, 2D motion of electrons.

The scheme of numerical simulation is as follows. In the plane geometry, the electron flow is represented as a set of large particles (charged leaves). For each of them, the relativistic equation

$$\frac{d^2 x_i}{dt^2} = -E(x_i), \quad (1)$$

is solved, where $x_i$ is the coordinate of an $i$th leaf, $E(x_i) = \partial\varphi/\partial x|_{x_i}$ is the space charge field strength at the point with coordinate $x_i$, and $\varphi$ is the space charge field potential.

Potential $\varphi$, field strength $E$, electron density $\rho$, electron velocity $v$, coordinate $x$, and time $t$ used in Eq. (1) are introduced through respective dimensionless variables marked by the prime (hereafter, we will use the dimensional variables and the prime will be omitted),

$$\varphi = (v_0^2/\eta)\varphi', \quad E = (v_0^2/L\eta)E', \quad \rho = \rho_0\rho', \\ v = v_0 v', \quad x = Lx', \quad t = (L/v_0)t', \quad (2)$$

where $\eta$ is the electron charge; $v_0$ and $\rho_0$ are, respectively, the static (undisturbed) electron velocity and density; and $L$ is the transit space length.

The space charge field potential and charge density are calculated on a uniform spatial mesh with step $\Delta x$. In the quasi-static one-dimensional approximation, the space charge potential obeys the Poisson equation

$$\frac{\partial^2 \varphi}{\partial x^2} = \alpha^2 \rho(x). \quad (3)$$

where $\alpha = \omega_P L/v_0$ is the Pierce parameter [37]. Space charge field strength $E(x)$ was determined by numerically differentiating the potential obtained. Equation (3) should be complemented by relevant boundary conditions,

$$\varphi(x = 0) = \varphi_0, \quad \varphi(x = 1) = \varphi_0 - \Delta\varphi, \quad (4)$$

where $\varphi_0$ is the accelerating potential ($\varphi_0 = 1$ in our normalization) and $\Delta\varphi$ is the decelerating potential across the grids.

The space charge density was found by linear weighting of the particles (leaves) on a spatial mesh with a reduced noise (the particle-in-cell method) [36]. In this method, the space charge density at a $j$th node of the spatial mesh, i.e., at a point with coordinate $x_j = j\Delta x$, is expressed as

$$\rho(x_j) = \frac{1}{n_0}\sum_{i=1}^{N}\Theta(x_i - x_j), \quad (5)$$

where $x_i$ is the coordinate of an $i$th particle, $N$ is the total number of large particles, $n_0$ is the parameter of the computing scheme that equals the number of particles per cell in the undisturbed state, and

$$\Theta(x) = \begin{cases} 1 - |x|/\Delta x, & |x| < \Delta x, \\ 0, & |x| > \Delta x, \end{cases} \quad (6)$$

is a piecewise linear shape function that specifies weighting of a large particle on a spatial mesh with step $\Delta x$.

Parameters $N_m$ (the number of nodes in the spatial mesh) and $n_0$ (the number of particles per cell in the undisturbed state) were set equal to $N_m = 800$ and $n_0 = 24$ (the number of particles in the domain of calculation in the undisturbed state is then $N = 19\,200$). Such values of the parameters provide a desired accuracy in analyzing complicated nonlinear processes, including deterministic chaos, in our electron–plasma system [36, 38]. The equation of motion was solved using the second-order step-by-step scheme [36], and the Poisson equation was integrated with the method of error vector propagation [6].

Injection of an electron beam with an initial longitudinal velocity spread was simulated with a modified version of the method used in [35, 36] to reproduce particle initial distributions in plasma systems.

As electron velocity initial distributions, which were specified at the entrance to the diode space, we took distribution functions $f(v)$ found experimentally (see Section 2 and Fig. 2, in which such a distribution for $\Delta v/v_0 \approx 3\%$ is shown). Then, we constructed function

$f(v)$ that is the distribution averaged over all points of measurement $r/r_b$ in the beam's cross section.

For injection of particles with velocities distributed according to $f(v)$, we constructed the integral distribution function

$$F(v) = \frac{\int_0^v f(v')dv'}{\int_0^{v_{max}} f(v')dv'}, \quad (7)$$

where $F(v = 0) = 0$, $F(v = v_{max}) = 1$, $v_{max}$ is the maximal velocity of electrons injected, and

$$\frac{dF(v)}{dv} = f(v)\left[\int_0^{v_{max}} f(v')dv'\right]^{-1}. \quad (8)$$

Now, if we equate function $F(v_s)$ to some distribution of numbers $R_s$ ($R_s \in (0, 1)$), it is easy to check that the distribution of $v_s$ will correspond to $f(v)$. Then, choosing a set of numbers $R_s$ (in our case, a total of 30 numbers, $s = 0, \ldots, 29$, distributed from 0 to 1 were considered) and integrating (7) numerically in small steps, we assign velocities $v_s$ to injected particles, with velocities $v_s$ being determined from the equality $D(v_s) = R_s$.

Thus, with the method described above, one can preset any experimental electron velocity distribution at the entrance to the space of interaction (plane $z_0$ in Fig. 1a).

## 4. NUMERICAL SIMULATION OF PHYSICAL PROCESSES IN A MULTIPLE-VELOCITY BEAM WITH A VIRTUAL CATHODE

Consider the results of numerically simulating the effect of velocity spread on chaotic oscillations in a system with a VC for Pierce parameter $\alpha = 0.9$ and $\Delta\varphi = 0.46$, where $\Delta\varphi$ is the decelerating potential of the second grid. At such parameters, the single-velocity beam exhibits chaotic oscillations. As $\Delta\varphi$ increases, chaotic oscillations in the VC beam become more and more complicated.

Beam current power spectra $P(f)$ calculated near the VC at different electron initial velocity spreads $\Delta v/v_0$ are shown in Figs. 6a–6c. The results of numerical simulation support the experimental data: as the initial velocity spread increases, oscillations in the VC beam become more and more entangled; specifically, the noise pedestal grows (the spectrum becomes noisy), irregularity parameter $N$ decreases, and the frequency band of the chaotic signal expands.

The same behavior is illustrated in Fig. 7, where mean noise intensity $S$ in the beam is plotted against irregularity parameter $N$ normalized to the irregularity parameter at the zero velocity spread, $\Delta v = 0$, and oscil-lation frequency band $\Delta f/f$ is plotted against initial velocity spread $\Delta v/v_0$ in the beam injected.

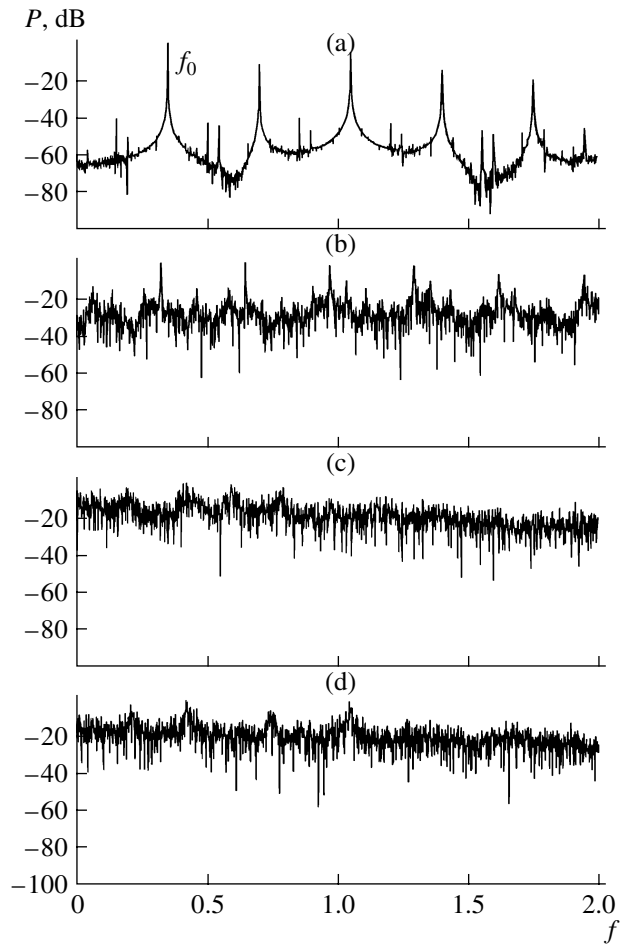

**Fig. 6.** Beam current power oscillation spectrum near the VC ($x = 0.94$) in the diode space with slowdown for Pierce parameter $\alpha = 0.9$ and different velocity spreads and decelerating potentials: (a) $\Delta v/v_0 = 0.5\%$, $\Delta\varphi = 0.46$; (b) $\Delta v/v_0 = 3\%$, $\Delta\varphi = 0.46$; (c) $\Delta v/v_0 = 5\%$, $\Delta\varphi = 0.46$; and (d) $\Delta v/v_0 = 3\%$, $\Delta\varphi = 0.6$.

From Figs. 6a and 7a, it follows that, when the velocity spread is small ($\Delta v/v_0 < 0.01$), the numerical simulation does not show any significant change in the system's behavior. For $\Delta v/v_0 = 0.01$–0.30, the chaotic oscillation frequency band expands, the noise spectral density $\log S$ increases, and the irregularity parameter decreases in the operating frequency band. However, when the velocity spread is high, $\Delta v/v_0 > 0.4$, the characteristics saturate at some level (i.e., cease to vary). Such a dynamics of the system with increasing velocity spread was also observed at other values of the Pierce parameter and decelerating potential of the second grid.

When the decelerating potential of the second grid grows with the velocity spread fixed, the dynamics of the system also becomes complicated. This is distinctly seen from the power spectra shown in Figs. 6b and 6d, which are constructed for the same velocity spread, $\Delta v/v_0 = 0.03$, and different decelerating potentials, $\Delta\varphi = 0.46$ and 0.60. It follows from these figures that the noise in the spectrum grows with decelerating potential: the basic spectral components and their harmonics arise on the background of the growing noise pedestal, which shades some of the harmonics in the oscillation spectrum. Simultaneously, the basic frequency of the power spectrum slightly shifts toward higher frequencies, since the VC oscillation frequency increases with decelerating potential [32].

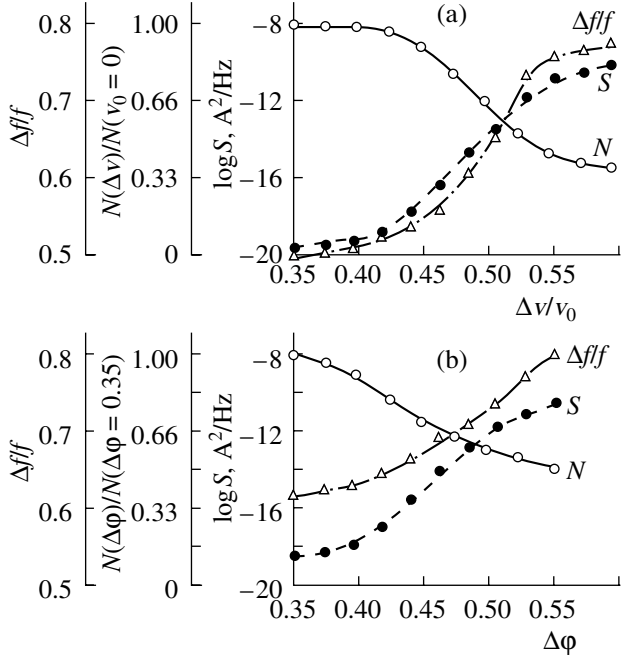

**Fig. 7.** Noise mean intensity $S$ in the beam, normalized spectrum irregularity $N$, and oscillation bandwidth $\Delta f/f$ vs. (a) electron initial velocity spread $\Delta v/v_0$ ($\alpha = 0.9$, $\Delta\varphi = 0.46$) and (b) decelerating potential $\Delta\varphi$ ($\alpha = 0.9$, $\Delta v/v_0 = 0.03$).

The related integral characteristics of the power spectra are shown in Fig. 7b, where they are plotted against the decelerating potential of the second grid. The range of the potential corresponds to chaotic oscillations in the VC beam for Pierce parameter $\alpha = 0.9$. The curves indicate that, in the presence of an initial velocity spread in the beam injected, an increase in the decelerating potential complicates the spectral characteristics of wide-band oscillations in the VC beam.

Let us concentrate on physical mechanisms responsible for the complicated dynamics of a VC electron beam at a high velocity spread. Here, numerical analysis of the system's model behavior is of key importance, since it allows for thoroughly considering physical processes in the space of interaction.

Following [17, 39, 40], we constructed various charged particle distribution functions in the space of interaction for small and large particle velocity spreads.

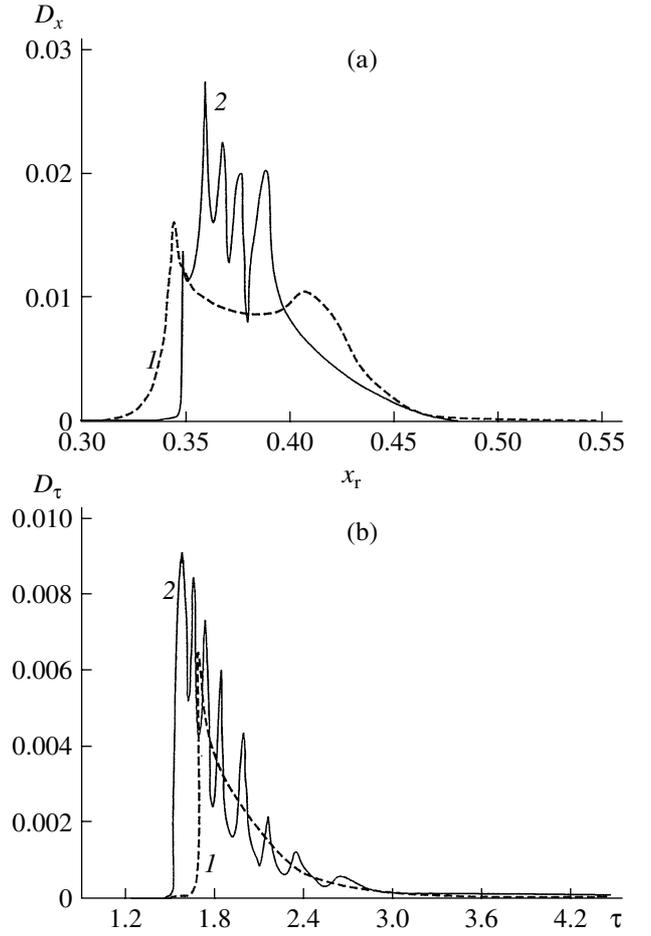

**Fig. 8.** (a) Distribution of the charged particle reflection planes (leaves) in the space of interaction and (b) charged particle lifetime distribution for the beam with *1*) small ($\Delta v/v_0 = 0.5\%$) and (*2*) large ($\Delta v/v_0 = 3\%$) velocity spread ($\alpha = 0.9$, $\Delta\varphi = 0.46$).

Figure 8 shows distributions $D_x$ of reflection positions $x_r$ of the charged particles and charged particle lifetime distributions $D_\tau$ ($\tau$ is the lifetime) in the space of interaction for the beam with small ($\Delta v/v_0 = 0.5\%$, dashed curve *1*) and large ($\Delta v/v_0 = 3.0\%$, continuous curve *2*) velocity spreads. From Fig. 8, it follows that the beam structure in the region where the particles reflect back toward the plane of injection (i.e., near the VC) changes qualitatively as the initial velocity spread increases.

When the spread is small, particle lifetime distribution function $D_\tau$ has a single maximum, which corresponds to the dynamics of a single structure forming in the system, a virtual cathode. Time $\tau_0$ at which the distribution function reaches the maximum, $D_{max} = D(\tau_0)$, is close to the characteristic time scale of oscillations of this single VC; accordingly, quantity $1/\tau_0$ is close to the frequency of the major spectral component in the power spectrum observed under the given conditions ($f_0$ in

Fig. 6a). Also, the distribution of the charged particles over the coordinates where they reflect back to the plane of injection shows that, in the case of a near-single-velocity beam, there exists a domain in the space of interaction, $x \in (0.32, 0.43)$, where the reflection probability of particles at each point of this domain is nearly the same (except for the extremities of this domain, where the distribution function exhibits small peaks). This domain may be arbitrarily referred to as the VC width, within which the VC oscillates. Thus, under small-spread conditions, a single basic electronic structure, virtual cathode, emerges in the system.

Under the large-spread conditions, the situation changes. In this case, the distribution functions have several distinct peaks. Such behavior of the multiple-velocity beam may be associated with the formation of several VCs (several space–time structures) at different distances from the plane of injection. This supposition is supported by the distribution of coordinates $x_r$ (Fig. 8), which are the points of reflection of the charged particles. This distribution function is heavily irregular and can be divided into four clear-cut subranges where the electrons reflect toward the plane of injection most frequently (i.e., has four peaks). Each of the peaks may be assigned to a VC localized within its own domain in the space of interaction. Note that the peaks lie on high bases, because each of the VCs oscillates in both space and time; however, the most probable localization sites of these structures are precisely the distribution peaks. The particle lifetime distribution is also highly irregular, which suggests that several electronic structures emerge in the multiple-velocity beam.

Each of these structures (VCs) has its own characteristic scale of oscillations. The partial reflection of the electron beam from each of the structures affects the formation conditions for other structures in the beam; accordingly, several internal feedback loops with different delay times arise. An increase in the velocity spread in the beam complicates the distributed feedback in the system, increasing the number of forming electronic structures. Thus, a high electron initial velocity spread in the beam adds to the chaotization of the system's dynamics, which shows up as the complication of the output radiation spectral characteristics.

## 5. DISCUSSION

Our investigation demonstrates that an increase in the electron velocity spread in the beam adds to the chaotization of VC oscillations. This shows up in the expansion of the oscillation spectrum, an increase in the noise spectral intensity, and a decrease in the irregularity of the spectrum in the operating frequency range. The experimental data qualitatively agree with the results of numerical simulation performed in terms of the diode model with slowdown. Applying such an approach, we gained a deeper insight into physical processes taking place in a multiple-velocity beam with a VC. The most significant result is that the additional chaotization of VC oscillations is due to the well-known mechanism [23, 32], namely, the formation of additional electronic structures, which were visualized by constructing distribution functions for charged particles.

However, compared with the experimental data, the numerical simulation in terms of the 1D model of electron beam dynamics gives a narrower oscillation bandwidth and a much higher irregularity of the spectrum. In our opinion, this is because the 1D theory was used, which ignores the basically 2D dynamics of electrons in the VC region. In particular, Fig. 6 shows that an increase in the electron longitudinal velocity spread goes in parallel with the extension of the distribution of electrons over angles of entry $\alpha$—the effect, which is taken into account in the 2D analysis of electron motion near a VC. It appears that elaborating a 2D theory of motion of an electron beam with a VC will be a mainstream direction in research on electron wave generators of wide-band chaotic signals.

In closing, the complication of wide-band noise-like oscillations in a VC beam (the fact predicted theoretically and verified experimentally) with increasing initial spread of electron velocities and angles of entry could be used for optimizing the parameters of controlled VC-based microwave generators of wide-band medium- and high-power signals [27, 28, 31, 32]

## ACKNOWLEDGMENTS

The authors thank Prof. D.I. Trubetskov for interest in this work, fruitful discussions, and valuable remarks.

This work was supported by the Russian Foundation for Basic Research (grant nos. 05-02-16286 and 05-02-08030), CRDF (grant no. REC-006), "Dynasty" federal scientific program, and International Center for Fundamental Physics in Moscow.